\documentclass[twocolumn,showpacs,preprintnumbers,amsmath,amssymb]{revtex4}

\usepackage{bm}
\usepackage{epsf}
\usepackage{dcolumn}

\def\threej#1#2#3#4#5#6{\left(\begin{array}{ccc}#1&#2&#3\\#4&#5&#6\end{array}\right)}
\def\ket#1{\left|#1\right\rangle}
\def\bra#1{\left\langle#1\right|}

\begin{document}

\widetext{

%

\title{On the recombination in high-order harmonic generation 
in molecules}

\author{B. Zimmermann, M. Lein, and J. M. Rost}

\affiliation{Max Planck Institute for the Physics of Complex Systems, 
N\"othnitzer Stra\ss{}e 38, 01187 Dresden, Germany}

\date{\today}

\begin{abstract}
We show that the dependence of high-order harmonic 
generation (HHG) on the molecular orientation 
can be understood within a theoretical treatment that does 
not involve the strong field of the laser. The results for H$_2$ show 
excellent agreement with time-dependent strong field calculations for 
model molecules, and this motivates a prediction for the orientation 
dependence of HHG from the N$_2$ 3$\sigma_g$ valence orbital. 
For both molecules, we find that the polarization of recombination photons 
is influenced by the molecular orientation.
The variations are particularly pronounced for the N$_2$ valence orbital,
which can 
be explained by the presence of atomic p-orbitals.
\end{abstract}


\maketitle
}

\section{Introduction}

In this paper we establish a connection  between photoionization/recombination 
(weak field processes) and high-order harmonic generation (a non-linear 
phenomenon in strong-field physics). The latter process is 
one of the most studied aspects of intense-laser physics 
because it serves
as a source of coherent radiation at high frequencies
\cite{krausz01,hentschel01}. 
High-order harmonic generation can be
explained by a recollision mechanism \cite{corkum93}: Close to
the maximum of the electric field of a femtosecond optical laser pulse 
a molecule is ionized. A free electron wave packet 
enters the continuum and follows the electric field of the
laser. If the laser is linearly polarized the electron will
approach the molecule again. The most energetic recollisions take
place near the second zero of the laser electric
field after electron release \cite{corkum93}. Hence, the
laser field at the time of recollision can be
considered as small. The optical laser drives the electronic wave 
packet far away from the molecule (as compared to the size of the molecule).
Due to rapid wave packet spreading the electronic wave packet will 
recollide approximately as a plane
wave with its momentum parallel to the laser polarization. 
Possible consequences of recollision are recombination, elastic
scattering or double ionization. In the recombination process
a photon is emitted, preferably parallel to the beam axis of 
the incident laser and with a frequency that is a multiple
of the incident laser frequency, therefore called high harmonic. 
By appropriate superposition of different harmonics one can create 
attosecond pulses which may be utilized to probe fast atomic and molecular
processes \cite{krausz01}. 
If the return time of the electronic wave packet
is well defined one can even think of using HHG itself
as a probe for time-dependent processes.
The ionization by the laser 
would represent the pump pulse and the recolliding wave packet 
would represent the probe pulse. The time between those two events
is shorter than an optical cycle of the laser.
This may open the door 
to the time-resolved investigation of
very fast atomic and molecular processes, cf. also the method described
in Ref. \cite{niikura03}. 


In recent years there has been growing interest in HHG from molecules. 
The dependence on molecular orientation has been studied experimentally 
\cite{hhgexp} and theoretically \cite{hhgtheo}. Considering the complexity 
of this process, 
theoretical investigations have been carried out mostly for H$_2$ 
and H$_2^+$ until now. 
How to overcome this?
As indicated above, at the time of recollision, when the radiative
recombination occurs, the electric field of the laser can be
considered to be small for the highest harmonics. In the following 
we will use an approximation in which the influence of the laser field 
on the recombination is considered to be even negligible
so that the computational methods developed in the context of
photoionization can be used.
Although this will not cover all the dynamics of HHG, it should
explain quite well dependencies of the high harmonics on the
molecular geometry and orientation. As we will show, this is indeed the
case. As a consequence it will be possible to describe HHG in much more
complicated targets in the future by shifting the focus
from the exact treatment of the time evolution 
towards the exact treatment of the final molecular interaction, the
recombination in high-order harmonic generation.

\section{Method}
In the recombination process the electron approaches the
molecular core and a photon is ejected, leaving the molecule predominantly 
in its ground state. The dynamics of this process is contained in its
transition amplitude. Since recombination is (microscopically) time
reversed photoionization one can use as recombination transition
amplitudes the complex conjugated photoionization transition
amplitudes. Furthermore, since we are here not interested in near threshold
behavior we can calculate those photoionization transition
amplitudes easily using the frozen core Hartree Fock (FCHF) method : The
molecular ground state wave function is derived in a
self-consistent-field approximation. The state of the ionized molecule 
is then obtained by removing one electronic charge out of the orbital
that is ionized. The molecule is not allowed to relax ('frozen
core'). The photoelectron orbitals were obtained using an
iterative procedure to solve the Lippmann-Schwinger equation
associated with the one-electron Schr\"odinger equation that
these orbitals satisfy (for further details see
\cite{lucchese82}).

For the mathematical description of the process the density matrix 
formalism \cite{blum81} has been applied (see also \cite{bzdr,lohmann}).
To be able do so we must model the recombination process.
We know the ground states of the neutral molecule and of 
the singly charged molecular core, 
$\ket{\Lambda_0}$ 
and $\ket{\Lambda_c}$, respectively. 
We assume that the molecular orientation
${\rm \bf m}$ does not change during the process. We also know the
state $\ket{{\rm \bf k}^{(+)}_e}$ of the incident electron.  
(In the following we will sum over not resolved molecular vibrational 
states and not resolved spin polarization states of the electron.)
Naturally, the photon, with a frequency $\omega_p$, will be polarized.
Therefore the full density matrix $\rho$ of the state 
after recombination in dipole approximation reads
\begin{eqnarray}
\rho &=& 
\ket{{\rm \bf m} \, \Lambda_0 \, \bm{\varepsilon} \, \omega_p} 
\bra{{\rm \bf m} \, \Lambda_0 \, \bm{\varepsilon} \, \omega_p}
\nonumber\\
&=& T \ket{{\rm \bf m} \, \Lambda_c \, {\rm \bf k}^{(+)}_e} 
\bra{{\rm \bf m} \, \Lambda_c \, {\rm \bf k}^{(+)}_e} T^+ ,
\end{eqnarray}
where $\bm{\varepsilon}$ is the polarization vector of the photon and
T the transition operator, i.e., the dipole operator. 
The photon properties will be measured in a detector in a 
direction ${\rm \bf n}_p$. For a perfect detector one gets by projecting on 
the different polarization states, which are   
$\ket{\lambda} \epsilon \{\ket{-1},\ket{0},\ket{1}\}$ 
in an arbitrary reference frame, for the matrix elements 
$\rho(\lambda,\lambda')$ of the density matrix $\rho$
\begin{equation}
\rho(\lambda, \lambda') =  
\bra{{\rm \bf m} \, \Lambda_0 \, {\rm \bf n}_p \lambda \, \omega_p} \rho
\ket{{\rm \bf m} \, \Lambda_0 \, {\rm \bf n}_p \, \lambda'  \, \omega_p}. 
\end{equation}
A common description of photon polarization employs the Stokes parameters.
The Stokes parameters are defined in a reference frame which z-axis 
is parallel to the photon momentum \cite{blum81}. In this frame the 
right (left) circularly polarized photon state is $\ket{+1}$ ($\ket{-1}$). 
($\ket{0}$ does not exist in this reference frame due to the transversal 
nature of the light.) The four Stokes parameters are : the total 
intensity $I$, the degree of circular
polarization $p_3 = (I_{+1} - I_{-1}) / I$ and the two degrees of linear
polarization $p_1 = (I(0^{\rm o}) - I(90^{\rm o})) / I$ and 
$p_2 = (I(45^{\rm o}) - I(135^{\rm o}))/I$. ($\varphi$ in $I(\varphi)$ starts 
at the x-axis in the xy-plane.) In the reference frame of the Stokes parameters 
one gets
\begin{subequations}
\begin{eqnarray}
I &=& \rho(1,1) + \rho(-1,-1)
\\
p_3 &=& \left( \rho(1,1) - \rho(-1,-1) \right) \cdot I^{-1}
\\
p_1 &=& - \left( \rho(1,-1) + \rho(-1,1)\right) \cdot I^{-1} 
\\
p_2 &=& -i \, \left( \rho(1,-1) - \rho(-1,1)\right) \cdot I^{-1}.
\end{eqnarray}
\end{subequations}
The electronic wave function  $\ket{{\rm \bf k}^{(+)}_e}$ can be expanded 
into spherical harmonics \cite{dill76}. However, due to the non-spherical 
molecular potential the dipole selection rules do not restrict the
expansion as in atoms. Nevertheless, both, bound and continuum 
electron wave functions converge quite rapidly. Therefore, to a very good
approximation a limited number of terms is sufficient, truncating the
expansion at a certain $l_{max}$. 
$\rho(\lambda, \lambda')$ 
can be split into a kinematical and a
geometrical part
\begin{widetext}
\begin{equation}
\rho(\lambda, \lambda')
= \sum_{\Gamma=0}^2 \sum_{{\cal L} = 0}^{2 \, l_{\rm max}} \sum_{d=|\Gamma -
{\cal L}|}^{\Gamma + {\cal L}}
M_{d {\cal L} \Gamma} 
\sum_{\gamma = -\Gamma}^{\Gamma}
(-)^{1-\lambda'} \left( 1 \lambda, 1 -\lambda' | \Gamma \gamma\right) 
\, Y^{d {\cal L}}_{\Gamma -\gamma}
\left( {\rm \bf m}, {\rm \bf k}^0_e\right). 
\label{rholls}
\end{equation}
where $(.,.|.)$ are Clebsch-Gordan coefficients. 
In Equ. \ref{rholls} we have used $l_{max}=10$.
The geometrical dependencies are expressed in bipolar spherical harmonics,
\begin{equation}
Y^{d {\cal L}}_{\Gamma -\gamma} \left( {\rm \bf m}, {\rm \bf k}^0_e \right) 
= \sum_{\delta=-d}^d \sum_{{\cal M} = - {\cal L}}^{\cal L}  
\left( d \delta, {\cal L} {\cal M} | \Gamma -\gamma\right) \, 
Y_{d \delta}\left( {\rm \bf m} \right) \,
Y_{{\cal L} {\cal M}}\left( {\rm \bf k}^0_e \right),
\end{equation}
with spherical harmonics $Y_{kq}$ and ${\rm \bf k}^0_e$ as the normalized  
electron momentum. The reference frame is given through the photon. 
The dynamical coefficient is 
\begin{eqnarray}
M_{d {\cal L} \Gamma} &=& \hat d \, \hat {\cal L}
\sum_{l=0}^{l_{\rm max}} \sum_{m_m = -(\Lambda^0+1)}^{\Lambda^0+1} 
\sum_{\lambda_m = -1}^1 
\sum_{l'=0}^{l_{\rm max}} \sum_{m'_m = -(\Lambda^0+1)}^{\Lambda^0 + 1} 
\sum_{\lambda'_m = -1}^1 
(-)^{1+m_m + \lambda'_m + \Gamma} \,  \hat l \, \hat l' \,  
R_{l m_m \lambda_m} \, R_{l' m'_m \lambda'_m}^*  
\nonumber \\
&& \times
\threej l {l'} {\cal L} 0 0 0
\threej l {l'} {\cal L} {-m_m} {m'_m} {\alpha_m} 
\threej 1 1 \Gamma {\lambda_m} {-\lambda'_m} {-\alpha_m}
\threej {\cal L} \Gamma d {\alpha_m} {-\alpha_m} 0 , 
\label{coefkin}
\end{eqnarray}
\end{widetext}
where $\Lambda^0$ is $0$ if the recombined orbital has 
$\sigma$ symmetry and $1$ if it has $\pi$ symmetry.
In Eq. (\ref{coefkin}) Wigner 3J symbols have been used. A hat over a 
quantum number means $\hat l = \sqrt{2 l + 1}$. The dynamical part is
calculated in the molecular body frame (symbolized by a sub-index $m$ 
at the quantum numbers).
By applying microscopic time reversal, the recombination matrix element 
in the molecular body frame in length form is 
\begin{equation}
R_{l m_m \lambda_m} \equiv \omega_p \left( i^{-l} \, 
\exp\left( i \Delta_l \right)
\bra{ \Lambda_c \, l m_m} d_{\lambda_m} \ket{ \Lambda_0 \lambda_m}\right)^* ,
\end{equation}
where $\omega_p$ is the energy of the photon,
$\Delta_l$ is the Coulomb phase shift, and $\bra{ \Lambda_c \, l m_m}
d_{\lambda_m} \ket{ \Lambda_0 \lambda_m}$ is the photoionization dipole 
matrix element in the body frame with the dipole operator $d_{\lambda_m}$.

\section{Results for H$_2$}

\begin{figure}[ht]
\epsfxsize 0.7\hsize
\centerline{\mbox{\epsfbox{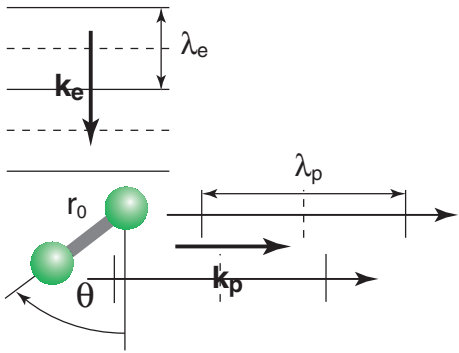}}}
\caption{The recombination geometry is shown schematically. The molecule with
a bond length $r_0$ is oriented relative 
to the electron momentum ${\rm \bf k}_e$ under 
an angle $\theta$. The photon is 
emitted
perpendicular to the electron
momentum.}
\label{figure1}
\end{figure}

\begin{figure}[ht]
\epsfxsize 1\hsize
\centerline{\mbox{\epsfbox{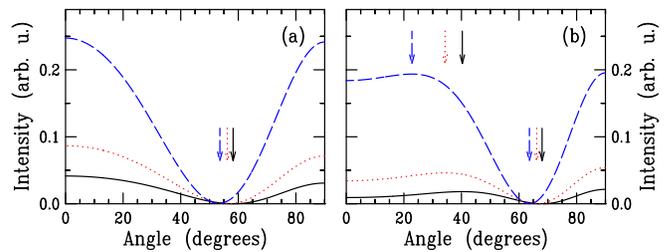}}}
\caption{
Dependence 
of the recombination photon intensity 
on the angle between molecular orientation and the electron momentum
for H$_2$. The molecule lies in the plane spanned by electron momentum
and photon direction 
[geometry (I)].
The solid curves are for an electron wavelength 
$\lambda_e$ of $1.4 \,\, \rm{a.u.}$, the dotted curves for 
$\lambda_e = 1.6 \,\, \rm{a.u.}$  and the dashed curves for 
$\lambda_e = 1.8 \,\, \rm{a.u.}$ (a) Molecular bond length 
$r_0 = 1.4 \,\, \rm{a.u.}$; (b) $r_0 = 2.0 \,\, \rm{a.u.}$. 
The positions of minima and maxima are marked by arrows. Obviously,
the positions of the extrema depend on the bond length and on the
wavelength of the electron. This behavior can be explained
by a two-center interference model (see text). 
Clearly, the minima are very pronounced.
}
\label{figure2}
\end{figure}

Using the formulae of the last section one can calculate the photon 
intensity and polarization, the Stokes parameters, as a function of the 
electron energy and for different orientations of the molecule and 
electron and photon propagation directions. 
Here we will focus on the HHG geometry (see Fig.
\ref{figure1}), where  the electron momentum is perpendicular to the 
emission direction of the photon. 
In the following, we distinguish between two cases:
(I) the molecule lies in the plane spanned by the 
electron momentum and the photon direction, and (II) the molecule
rotates in the plane perpendicular to the photon direction.  
We have first calculated the intensity for H$_2$ 
recombination in geometry (I) as a function of the molecular orientation.
Here, the photon is polarized parallel to the electron
momentum for symmetry reasons.
The calculation was carried out for different bond lengths and for 
different wavelengths of the  electron (see Fig. \ref{figure2}). 
One finds a pronounced minimum, which shifts if one changes 
the  bond length of the molecule or if one changes the wavelength of the 
electron.

One can explain the general behavior at electron wavelengths comparable 
to the bond length of the molecule by the well-known two-center 
interference. Here one imagines the diatomic molecule as two centers 
which are hit coherently by the same plane electron wave, but with a
phase difference that depends on 
the molecular orientation towards the electron. 
Recombination leads to the ejection of a photon. Interference occurs, 
since it is not known which center has emitted the photon. Changing 
the electron wavelength and/or the molecular orientation will alter the 
phase difference so that an interference pattern will be obtained. At the 
energies used here, the photon wavelength is much larger than the 
dimension of the molecule, so that one can neglect the phase 
shift resulting from the orientation of the molecule with respect to 
the photon. The bond length $r_0$ of the molecule, the electron 
wavelength $\lambda_e$ and the angle $\theta_{ex}$ between molecular 
axis
and propagation direction of the electron under which 
extrema in the recombination photon intensity appear 
(see Fig. \ref{figure1}) are then - in the two-center interference 
picture - related through
\begin{equation}
r_0 \, \cos (\theta_{ex}) = \frac{n}{2}\, \lambda_e + \delta_\theta,
\qquad  n = 0,1,2,\dots
\label{twocentereq}
\end{equation}
where $\delta_\theta$ is the difference of additional phase shifts the 
electronic wave function experiences in the vicinity of the nuclei. In 
the ideal case those phase shifts are equal and $\delta_\theta$ is zero.
$\delta_\theta$ depends on the orientation of the molecule 
and is expected to be maximal 
if the molecule is parallel to the electron momentum 
and zero if perpendicular. If $\delta_\theta$ is small,
interference will be constructive for even $n$ in Eq. 
(\ref{twocentereq}) and destructive for odd $n$. Parallel and 
perpendicular orientation of the molecule relative to the electron momentum 
always give rise to trivial extrema. For fixed bond length $r_0$
and increasing electron wavelength, minima will occur at 
positions where the molecule is more and more aligned along the electron 
momentum, up to the point where both are parallel.  In the process the 
minimum gets less 
pronounced and its absolute value is not
approximately zero anymore.


\begin{figure}[ht]
\epsfxsize 0.8\hsize
\centerline{\mbox{\epsfbox{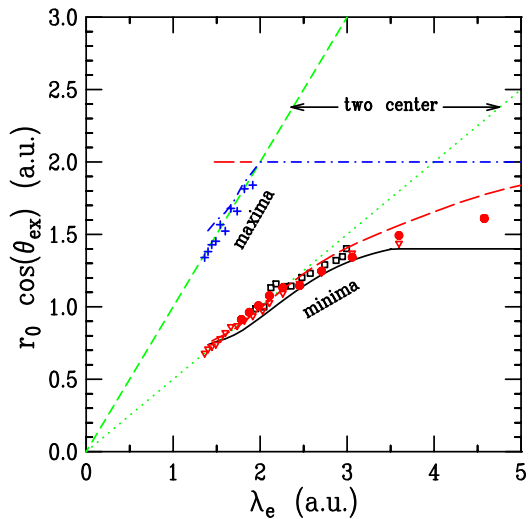}}}
\caption{
Recombination in geometry (I) for H$_2$.
For the extrema in the photon-intensity orientation dependence,
the projection 
of the bond length on the electron momentum direction,
$r_0\cos(\theta_{ex})$, is plotted against the electron 
wavelength $\lambda_e$.
The dotted (dashed) curve marks minima (maxima) due to 
two-centre emittance (see text). 
The solid curve represents minima for 
$r_0= 1.4 \,\, {\rm a.u.}$, whereas the 
long-dashed (dashed-dotted) curves represent minima (maxima) for 
$r_0= 2 \,\, {\rm a.u.}$. For comparison the extrema in the orientation 
dependence in HHG for time-dependent strong field calculations for 
model molecules \cite{lein02} are plotted as well: interference minima for 
($\triangledown$) H$^+_2$ at $r_0 = 2 \,\, {\rm a.u.}$, 
laser intensity $I=10^{15} \, {\rm W}/{\rm cm}^2$; 
($\bullet$) H$^+_2$ at $r_0= 2\,\, {\rm a.u.}$, 
$I=5\times 10^{14} \, {\rm W} / {\rm cm}^2$; 
($\square$) H$_2$ at $r_0 = 1.4 \,\, {\rm a.u.}$, 
$I=5\times 10^{14} \, {\rm W} / {\rm cm}^2$;
($+$) interference maxima for H$^+_2$ at $r_0= 2 \,\, {\rm a.u.}$,
$I=10^{15} \, {\rm W}/ {\rm cm}^2$. Note that the present definition of
$\lambda_e$ differs from \cite{lein02}.}
\label{figure3}
\end{figure}

A convenient way of analyzing the extrema is presented in Fig.~\ref{figure3}
where the projection $r_0\cos(\theta_{ex})$ is plotted as a function
of $\lambda_e$.
Our results bear a 
strong resemblance to those of time-dependent strong field calculations 
of HHG in H$_2$ and H$^+_2$ model molecules \cite{lein02}. 
This supports our prior assumption on that one can treat
the recombination in HHG approximately as a weak-field process.
Also, one finds the predictions made about the increase of 
$\delta_\theta$ towards parallel molecular orientation confirmed.
Not surprisingly, the signature of two-center interference 
fades with increasing electron wavelength. However, at the electron 
wavelengths
considered, this  effect can be attributed mainly to 
the decreasing ratio of kinetic energy of the electron to the ionization
threshold. In general, the orientation dependence of the recombination photon 
intensity for H$_2$ can be  well described within the two-center 
interference model.

\begin{figure}
\epsfxsize 0.9\hsize
\centerline{\mbox{\epsfbox{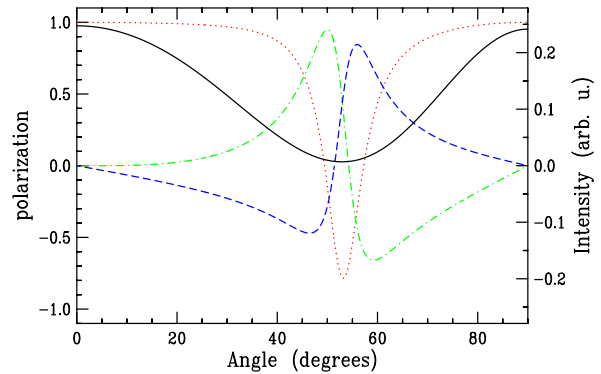}}}
\caption{The H$_2$ molecule lies in the plane perpendicular to 
the photon propagation direction which is parallel to the z-axis in the
Stokes parameters frame
[geometry (II)].
The electron is here chosen to move
along the x-axis.  Shown is as solid curve the recombination photon 
intensity for $\lambda_e = 1.6 \,\, {\rm a.u.}$, 
$r_0 = 1.4 \,\, {\rm a.u.}$ against the angle between the molecular axis
and the propagation 
direction of the electron. Dotted, dashed-dotted 
and dashed curves are the Stokes parameters $p_1$, $p_2$ and $p_3$, 
respectively. 
These show strong 
variations in the vicinity of the minima.
}
\label{figure4}
\end{figure}


As indicated, we can calculate all the Stokes parameters. 
In geometry (I), 
only linear photon polarization is possible. In geometry (II), however, 
where the molecular axis lies in the plane perpendicular to the photon 
propagation direction, the photon can have different polarizations and 
even circular polarization can be obtained (see Fig. \ref{figure4}). 
All polarizations show strong variations in the vicinity of the 
interference minimum. Otherwise only small polarization 
variations have been found.
Thus, although
the polarization depends on the geometry,
the difference is small for H$_2$
because the signal is dominated by the polarization parallel to the electron
momentum except in the small range around the interference minimum.
This can be understood within the two-center model since the H$_2$
molecular orbital is approximately the sum of two atomic 1s-orbitals. These
are spherically symmetric and therefore do not produce a signal polarized 
perpendicular to the electron momentum.

\section{Results for N$_2$}

Given the excellent agreement of our H$_2$ results with time-dependent strong
field calculations \cite{lein02} we can move to a prediction for the 
orientation dependence of HHG from the N$_2$ $3\sigma_g$ valence orbital.
The time-dependent HHG calculation for this system is quite complicated 
and has not been carried out.

\begin{figure}[ht]
\epsfxsize 1\hsize
\centerline{\mbox{\epsfbox{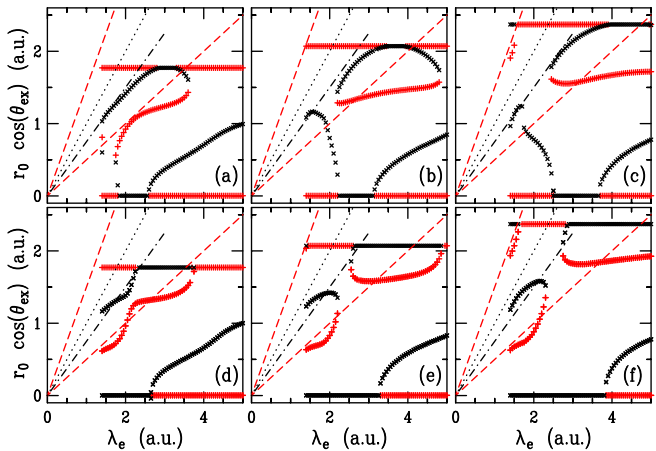}}}
\caption{
Projection $r_0\cos(\theta_{ex})$ 
versus electron wavelength $\lambda_e$ for molecular orientations
$\theta_{ex}$ under which
minima (x) and maxima (+) occur in the N$_2$ $3\sigma_g$ 
recombination photon intensity.
The bond length $r_0$ is $1.768 \,\, {\rm a.u.}$ in (a) and (d), 
$2.068 \,\, {\rm a.u.}$ in (b) and (e)
(ground-state bond length),
and $2.368 \,\, {\rm a.u.}$ in (c) and (f), respectively.
The upper plots [(a), (b), (c)] show the positions of extrema when the
molecule lies in the plane spanned by the electron and photon direction
[geometry (I)],
whereas the molecule rotates in the plane perpendicular to the photon
direction in the lower plots (d), (e), (f) [geometry (II)].
The dotted (dashed) curve is the one for minima (maxima) according to the
two-center interference model for p-orbital contributions.
Note that minima and maxima are interchanged as compared to the s-orbital 
contributions. The dashed-dotted line marks minima according to 
$r_0\cos(\theta_{ex})=0.75 \,\lambda_e$.}
\label{figure5}
\end{figure}

\begin{figure}[ht]
\epsfxsize 0.9\hsize
\centerline{\mbox{\epsfbox{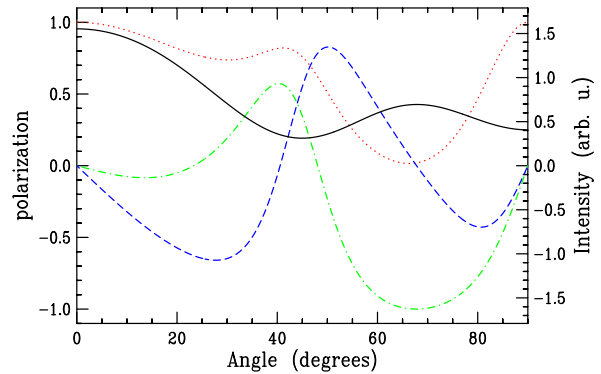}}}
\caption{The N$_2$ molecule lies in the plane perpendicular to 
the photon propagation direction which is parallel to the z-axis in the
Stokes parameters frame
[geometry (II)].
The electron is here chosen to move
along the x-axis.  Shown is as solid curve the recombination photon 
intensity for $\lambda_e = 1.6 \,\, {\rm a.u.}$,  
$r_0 = 2.068 \,\, {\rm a.u.}$ against the angle between the molecular axis
and the propagation direction of the electron. 
Dotted, dashed-dotted 
and dashed curves are the Stokes parameters $p_1$, $p_2$ and $p_3$, 
respectively.}
\label{figure6}
\end{figure}

While both, H$_2$ $1\sigma_g$ and N$_2$ $3\sigma_g$, have the same symmetry, 
they are rather different otherwise.
While $1\sigma_g$ is mainly built up from 
atomic s-orbitals and does not possess nodes,
$3\sigma_g$ is dominated by atomic p-orbitals and has a more complex 
structure \cite{wahl66}. As a consequence the orientation dependence for 
N$_2$ $3\sigma_g$ is more complex than  for H$_2$ $1\sigma_g$.
As in the previous section,
we have investigated geometries (I) and (II).
In Fig. \ref{figure5} the extrema for the equilibrium 
bond length $2.068 \, {\rm a.u.}$ as well as for $1.768 \, {\rm a.u.}$ 
and $2.368 \, {\rm a.u.}$ are plotted. 
Contrary to H$_2$, there are big differences between the two geometries. 
Figure \ref{figure6} shows the orientation 
dependence 
of the Stokes parameters
for geometry (II). Large variations are found over a broader range of angles
than in H$_2$, i.e., the component perpendicular to the electron
momentum cannot be disregarded. At small angles, the signal is still
dominated by the polarization parallel to the electron, but not so 
for larger angles.

Since N$_2$ is a  homonuclear diatomic molecule we might expect to 
find two-center interference 
in the region where the wavelength of the electron equals
approximately the internuclear distance. However, Fig. \ref{figure5}
shows that it is not straightforward to identify such signatures.

To understand the observed behavior, we first note that the 
two atomic p-orbitals ``inside'' the N$_2$ valence orbital are of course
not spherically symmetric. Therefore, unlike s-orbitals, 
each of them can produce a substantial component
polarized perpendicular to the electron momentum
with a pronounced dependence on the molecular orientation. However, this
component 
does not show up 
in geometry (I). This explains the difference
between the two geometries. Furthermore, the molecular orbital
is not constructed of atomic p-orbitals only, but we have an s-orbital 
admixture of about 30\%. This makes the two-center
interference picture problematic because different interference patterns
are expected for the two orbital types: to ensure $\sigma_g$ symmetry,
the two s-orbitals 
$\phi_{\rm s}(\mathbf{r}-\mathbf{r}_0/2)$ and 
$\phi_{\rm s}(\mathbf{r}+\mathbf{r}_0/2)$
are added with the same sign, so that interference conditions
are obtained as described for H$_2$; the p-orbitals 
$\phi_{\rm p}(\mathbf{r}-\mathbf{r}_0/2)$ and
$\phi_{\rm p}(\mathbf{r}+\mathbf{r}_0/2)$, on the other hand, 
are added with opposite signs, leading to an interchange of maxima and minima
\cite{lein02}. 

\begin{figure}[ht]
\epsfxsize 1\hsize
\centerline{\mbox{\epsfbox{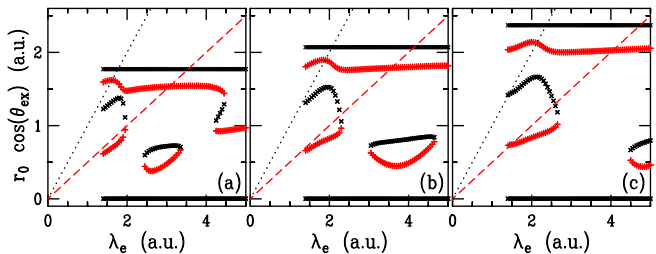}}}
\caption{Same as Fig. \ref{figure5} when only the component
polarized perpendicular to the electron momentum is measured 
in geometry (II). (The molecular axis lies in a plane perpendicular to the
the photon propagation direction.) The dotted (dashed) curve is the 
one for minima (maxima) according to the two-center interference model for
p-orbital contributions.}
\label{figure7}
\end{figure}

Clearly, the simultaneous presence of both types of
interference will lead to a complicated orientation dependence.
However, we can look for situations where either s- or p-orbitals
dominate the signal. One such case is geometry (I) near an orientation
of 90$^{\rm o}$. Here, the individual p-orbitals generate a 
negligible signal due to their mirror antisymmetry, as is obvious
when one considers a matrix element of the form 
$\langle\phi_{\rm p_x}(\mathbf{r})|z|\exp(ikz)\rangle$ where
the incoming electron is approximated by a plane wave. 
Consequently, s-orbital
contributions should dominate. In fact, Figs.~\ref{figure5}(a)-(c) show
that in the small-wavelength regime, geometry (I) yields
local maxima at $\theta=90^{\rm o}$  as a consequence of
constructive interference (similar to H$_2$).
Another example is geometry (II), when only the component perpendicular
to the electron momentum is measured. In this case, the s-orbital 
contributions are small as explained above in the context of H$_2$.
Consequently, we should observe the p-type interference pattern, i.e.,
zero signal at 90$^{\rm o}$ and a series of minima and maxima
when the angle is decreased. This is indeed the case for small electron
wavelengths as is shown in 
Fig. \ref{figure7} where the extrema for the perpendicular
component are plotted. In this plot, the extrema are systematically slightly 
below the ``perfect'' two-center interference lines 
$r_0\cos(\theta_{ex})=n\,\lambda_e/2$. The
total photon intensity in geometry (II) exhibits a similar behavior,
see the lower panels of Fig. \ref{figure5}. This demonstrates
the predominance of the p-orbital part even in the total signal.

In geometry (I) we have both s- and p-orbital contributions 
for the small and intermediate angles. 
Although in this regime
the results cannot be explained in a simplified picture, we find a
set of minima (see Fig. \ref{figure5}) 
following a straight line 
$r_0\cos(\theta_{ex})=0.75\, \lambda_e$ that lies just in the middle
between the two-center interference lines.

\section{Conclusions}

In conclusion, we have shown that the orientation dependence of the 
recombination photon intensity in H$_2$ can be described very well in a 
two-center interference model. Our results on the orientation dependence
bear a remarkable resemblance with those obtained from the time-dependent 
Schr\"odinger equation for HHG \cite{lein02}. This shows that our method 
can be used alongside those others to obtain estimates about the effects of
molecular geometry and orientation on the photon intensity in HHG.
We have made such a prediction for the case of N$_2$. Furthermore, we have
demonstrated, that the photons from HHG in oriented molecules do not 
exhibit only linear polarization. Rather, the polarization of those
photons show strong variations and can even be circular, depending on 
the molecular orientation.
For N$_2$, the interpretation of the results within
a two-center interference picture is hampered by the fact that 
the valence orbital has admixtures of both atomic s- and p-orbitals,
which produce different interference patterns. However, we have pointed
out situations where one of the two orbital types dominates the signal
so that interference can be observed.


\begin{thebibliography}{99}

\bibitem{krausz01}
F. Krausz, Phys. World {\bf 14}, 41 (2001).

\bibitem{hentschel01}
M. Hentschel, R. Kienberger, Ch. Spielmann, G. A. Reider, N. Milosevic,
T. Brabec, U. Heinzmann, M. Drescher, and F. Krausz, Nature {\bf 414},
509 (2001).

\bibitem{corkum93}
P. B. Corkum, Phys. Rev. Lett. {\bf 71}, 1994 (1993).

\bibitem{niikura03}
H. Niikura, F. L\'egar\'e, R. Hasbani, M. Yu Ivanov, D. M. Villeneuve,
and P. B. Corkum, Nature {\bf 421} (2003).

\bibitem{hhgexp} 
R. Velotta, N. Hay, M. B. Mason, M. Castillejo, and J. P. Marangos, 
Phys. Rev. Lett. {\bf 87}, 183901 (2001).

\bibitem{hhgtheo} 
H. Yu and A. D. Bandrauk, Chem. Phys. {\bf 102}, 1257 (1995); 
R. Kopold, W. Becker, and M. Kleber, Phys. Rev. A {\bf 58}, 4022 (1998);
D. G. Lappas and  J. P. Marangos, J. Phys. B : At. Mol. Opt. Phys. 
{\bf 33}, 4679 (2000).

\bibitem{lucchese82}
R. R. Lucchese, G. Raseev, and V. McKoy, Phys. Rev. A {\bf 25}, 2572 (1982).

\bibitem{blum81}
K. Blum,  {\it Density Matrix and Its Application} (Plenum Press, 
New York, 1981).

\bibitem{bzdr}
B. Zimmermann,  {\it Vollst\"andige Experimente in der atomaren und
molekularen Photoionisation}, vol 13 in  {\it Studies of Vacuum and X-ray 
Processes}  edited by U. Becker (Wissenschaft und Technik Verlag, Berlin, 2000).

\bibitem{lohmann} 
B. Lohmann, B. Zimmermann, H. Kleinpoppen, and U. Becker,
Adv. At. Mol. Opt. Phys. {\bf 49}, 218 (2003).

\bibitem{dill76}
D. Dill, J. Chem. Phys. {\bf 65}, 1130 (1976).

\bibitem{lein02}
M. Lein, N. Hay, R. Velotta, J. P. Marangos, and P. L. Knight, Phys. Rev. A
{\bf 66}, 023805 (2002).

\bibitem{wahl66}
A. C. Wahl, Science {\bf 151}, 961 (1966).

\end{thebibliography}
\end{document}